# Ideality factor in transport theory of Schottky barrier diodes

A.A.Klyukanov, P.A.Gashin, R.Scurtu
*Semiconductor Physics Lab, Faculty of Physics, Moldova State University, 60 A. Mateevici str., Chisinau MD 2009,Republic of Moldova,.*
klukanov@usm.md

*Abstract* — A microscopic many-body transport approach for electronic properties of spatially inhomogeneous systems is developed at the fully quantum-mechanical level by means of plane wavelets second quantization representation. It is obtained that current density is determined by the statistically averaged microscopic polarization, dependent on the quantized positions and quantized momenta of charge carriers. Distribution function of electrons includes many-body effects via drift, diffusion and thermionic emission as well as entirely quantum-mechanical tunneling through a Schottky barrier. Dependences of the current versus voltage on the thickness of semiconductor layer, the relaxation times in the neutral region and in the depletion layer, the width of Schottky barrier and the mean free paths are investigated. It is established that ideality factor $n$ is a function of applied voltage $V$ and depends on a relation between the width of Schottky barrier and depletion layer. The value of $V$ at which I-V characteristics acquire an ohmic nature is depended on the parameters of semiconductors.

## 1. Introduction.

Solar cells based on the thin film technology are important candidates for photovoltaic industry. One of significant techniques for the device characterization is current versus voltage. Forward current, in many cases, can be described by exponential dependence with exponent index $qV/nkT$, where n is the ideality factor lied in the range 1 to 2. Such simple formula does not cover the observed variety of experimental I-V characteristics of MOS and MS structures (frequently $n>2$.). The measured I-V characteristics of MS structures and their evolution at the temperature variation are governed by the thermionic-emission-diffusion Sze-Crowell theory [1-3]. Analysis of semiconductor-device operation and behavior of carriers under influences of applied electric field necessary therefore is described by the basic equations which include the Poisson equation, continuity equations and current-density equations. With account of drift and diffusion according to the both Ohm's law and Fick's law we have

$$j(x) = q\mu_n n(x)E(x) + \mu_n kT\frac{dn(x)}{dx}, \quad (1)$$

where $E(x)$ is the intensity of the field, n(x) is the density, $\mu_n$ is the mobility, j is the current density and q is the charge of electrons. Other designations are standard. It is well known that equation for current density (1) can easily be derived by using of solution of classical Boltzmann equation in the so called relaxation time $\tau(x,k_x)$ approximation

$$n(x,k_x)=n_e(x)n_e(k_x)g(x,k_x), n_e(k_x)=\exp\left(-\frac{\hbar^2k_x^2}{2mkT}\right)\left(\frac{\pi\hbar^2}{2mkT}\right)^{1/2}, g(x,k_x)=1-\frac{\hbar k_x}{m}\left(\frac{qE(x)}{kT}+\frac{n'(x)}{n(x)}\right)\tau(x,k_x) \quad (2)$$

Here m is the effective mass of electrons, $n_e(k_x)$ is the equilibrium distribution function. At small field intensities $n_e(x)$ is the thermodynamic equilibrium concentration of electrons and $g(x,k_x)\cong 1$. If one wants to investigate high field transport one may introduce the mobility depending on $E(x)$ [4]. We consider nonlinear current-voltage characteristics by means of concentration $n(x)$ as a function of arbitrary field intensity $E(x)$. Classical description (1,2) is based on the classical Boltzmann equation and phase space model according to which dynamical variables $x, p_x$ specify the state of a classical system. For quantum systems a simultaneous specification of coordinates $x_i$ and momenta $p_i = \hbar k_i$ is not possible in view of Heisenberg uncertainty relations. A quantum mechanical description consists of a Hilbert space of states. According to the correspondence principle, the laws of quantum mechanics must reduce to those of classic in the limit where $\hbar$ tends to zero. This fundamental requirement views the equations of classical mechanics as limit of the Schrödinger equation. So Boltzmann equation can be derived from the Schrödinger evolution of interacting particles. But classical description of homogeneous system is just the same as the quantum description if one uses a plane-waves representation. Thus some of the Boltzmann equation driving terms may be derived from the quantum mechanical many-body analysis for expectation value of microscopic polarization

$$P_{kk'}(t) = <\hat{P}_{kk'}(t)> = <a_k^+(t)\, a_{k'}(t)> \quad (3)$$

by making use a plane-waves representation in a fairly direct way. In such a manner account of particles interaction with homogeneous electric field $E$ leads to the drift term, represented in the form (1, 2). But it is much more difficult to take into consideration the dependence of distribution function on position of particles which can not be considered in a plane-waves representation. In order to develop kinetic theory with simultaneous listing of coordinates and momenta one has to introduce Wigner representation [5,6]. Wigner distribution function is derived from Greens function using the Wigner transform, which is Fourier transform, with respect to the relative coordinate. This technique is useful in the systems that are not homogeneous. Wigner distribution function is reduced to classic Boltzmann function in the limit where $\hbar$ tends to zero. But for many Hamiltonians of interest Wigner distribution function is not positive definite and hence can not be interpreted as a probability density. To construct the transport theory one has to put into the consideration a number of simplifying assumptions. Here we consider an implication of difference kinetic theory, suggested in the reference [7], which, in a sense, allows us to introduce a distribution function and consider of a validity of some theoretical assumptions. We discuss our analysis of kinetic equations for semiconductor-device operation in relation to other theoretical approaches.

## 2. Quantum mechanical problem analysis.

To arrange our proofs of solution for distribution function $n(X, K_x)$ we turn to account method of difference kinetic equations [7] considering discrete phase space $(X, K_x)$ by means of plane wavelets representation

$$|X, K_x\rangle = \Psi_{X,K_x}(x) = \frac{1}{\sqrt{d_x}} \exp(iK_x x)\, \theta_+(X + \frac{d_x}{2} - x)\theta_-(x - X + \frac{d_x}{2}),$$

$$K_x = \Delta K_x n_x, \Delta K_x = \frac{2\pi}{d_x}, X = d_x m_x, \quad \theta_-(x) = \begin{cases} 0, & x < 0 \\ 1, & x \ge 0 \end{cases}, \quad \theta_+(x) = \begin{cases} 0, & x \le 0 \\ 1, & x > 0 \end{cases} \quad (4)$$

Product of step functions $\theta_+\theta_-$ in Eq.(4) is the wavelet scaling function. Plane wavelet function $|X, K_x\rangle$ (4) is determined on the interval $[x - d_x/2, x + d_x/2)$. Position $X$ and momentum $K_x$ of electrons are quantized according to Eq. (4), where $n_x, m_x$-are integers $n_x = 0, \pm 1, \pm 2, \ldots \pm \infty$, $m_x = 0, \pm 1, \pm 2, \ldots \pm (d/2d_x - 1/2)$. where d is a width of a semiconductor. It was shown [7] that the set of plane wavelet orthonormal functions $|X, K_x\rangle$ (4) is complete and can be used as the second quantization basis allowing introducing the positively definite distribution function $n(X, K_x) = P_{kk}$ that can be considered as density of particles of inhomogeneous system described by numbers of particles at quantized positions $(X,)$ with quantized momenta. $(K_x)$. To introduce second quantization operators and determine current density $j_x(X)$ it is necessary to express the $j_x(X)$ via the classical sum over all charges and then enter second quantization operators

$$Sd_x\sum_X j_x(X) = -\frac{q}{m}\sum_i p_{xi} = -\frac{q}{m}\sum_{kk'}(\hat{p}_x)_{kk'} < a_k^+ a_{k'} >, \quad \hat{p}_x = -i\hbar\frac{\partial}{\partial x} \qquad (5)$$

where S-surface area of the semiconductor. It is now straightforward to insert moment operator $\hat{p}_x$ into the equation for $j_x(X)$ and evaluate matrix elements $(\hat{p}_x)_{kk'}$. Using plane wavelets representation (4) one gets for the current density the next general expression

$$j_x(X) = \frac{q\hbar}{mSd_x}\sum_{K_x}\{K_x P_{kk} + I_k\},\ I_k = \frac{i}{2d_x}\sum_{K_x'}(-1)^{\frac{K_x'-K_x}{\Delta K_x}}\left(P(K_x, X-d_x; K_x', X) - P(K_x, X; K_x', X-d_x)\right) \qquad (6)$$

without of any approximations. The first term on the right – hand side of the equation (6) has the well-known classical analog for current density. The second term $I_k$ is quantum mechanical probability flux and has no analog in classic theory. Indeed, if we use the limit where function $P_{kk'}$ varies little in $\Delta X = d_x$, and $\Delta K_x = 2\pi / d_x$ then expanding $P(X - d_x)$ around $d_x = 0$ equation (6) at $k = k'$ becomes

$$I_k = -\frac{i}{2}\left(P_k^*\frac{dP_k}{dx} - P_k\frac{dP_k^*}{dx}\right),\ P_{kk} = P_k^* P_k \qquad (7)$$

Distribution function $P_{kk} = P_k^* P_k$ corresponds in structure to density of probability. Thus the problem represented by Eq.(6) reduces to determination of the expectation value of microscopic polarization $P_{kk'}(t) = < \hat{P}_{kk'}(t) >$. Difference kinetic equation for microscopic polarization with account of two-particle correlations

$$\frac{\partial P_{kk'}(t)}{\partial t} = \frac{i}{\hbar} < [\hat{H}(t), \hat{P}_{kk'}(t)] > + \left(\frac{\partial P_{kk'}(t)}{\partial t}\right)_{sc} \qquad (8)$$

was derived quantum mechanically [7]. Here $\hat{H}(t)$ is the Hartree-Fock Hamiltonian in a plane wavelets representation (4). The general explicit expression of the Eq.(8) which is not given due to its cumbersome form [7], permits to evaluate the microscopic polarization $P_{kk'}(t)$ at arbitrary $k$ and $k'$. It was shown [7] that quantum difference kinetic equation (8) for distribution function at $k = k'$ transforms into the classical Boltzmann equation in the limit, where expectation value of particles number $P_{kk}$ varies little in $\Delta X = d_x$, and $\Delta K_x = 2\pi / d_x$. The last term in the Eq. (8) is from the scattering. It depends on the averages $P_{kk'}(t) = < a_k^+(t) a_{k'}(t) >$ and $< a_{k'}(t) a_k^+(t) >$. According to the simple decoupling scheme $< a_{k'}(t) a_k^+(t) > = < a_k^+(t) a_{k'}(t) > = < a_{k'}(t) >< a_k^+(t) >$ distribution function is expressed in the multiplicative form as $P_{kk} = P_k^* P_k$. Having used of the same approximation (the limit where $P_k$ varies little in $\Delta X = d_x$, and $\Delta K_x = 2\pi / d_x$) at $k = k'$ one gets

$$\frac{\partial P_k}{\partial t} = -\frac{i}{\hbar}\left(-\frac{\hbar^2}{2m}\frac{\partial^2}{\partial X^2} + E_k\right)P_k - \frac{q}{\hbar}E(X)\frac{\partial P_k}{\partial K_x} - \frac{\hbar}{m}K_x\frac{\partial P_k}{\partial X},\ E_k = \frac{\hbar^2 K_x^2}{2m} + U(X) + \Delta E_k \qquad (9)$$

Equation (9) combines both Schrödinger (imaginary part) and Boltzmann (real part) equations. Hence the solution for $P_{kk'}(t)$ is given by $P_{kk'}(t) = P_{kk}(t)\exp(i\varphi_k)\exp(-i\varphi_{k'})$, where $\varphi_k$ is the phase of state. With account of fluctuations of instantaneous field near the mean field we have [7]

$$\Delta E_k(t) = -\frac{i}{\pi}\sum_q V_q\int_{-\infty}^{\infty} d\omega[n(\omega)+1]\,\mathrm{Im}\left\{\frac{1}{\varepsilon^*(q,\omega)}\right\}\int_0^t\left(\sum_{k_1}\left|e_{kk_1}^{iqr}\right|^2 G_{k_1k}(t,t_1)\Phi_{k_1}^{\omega}(t-t_1)\right)dt_1 \qquad (10)$$

where $\varepsilon(q,\omega)$ is the dielectric function, $n(\omega) = \left[\exp(\hbar\omega / k_0 T) - 1\right]^{-1}$, $V_q$ is the Fourier transform of Coulomb potential. Functions $G_{k_1k}(t,t_1)$ and $\Phi_{k_1}^{\omega}(t-t_1)$ are determined as

$$\Phi_k^{\omega}(t-t_1)=n_k\exp(i\omega(t-t_1))+(1-n_k)\exp(-i\omega(t-t_1)), G_{k_1k}(t,t_1)=\exp\left(\frac{i}{\hbar}\int_{t_1}^{t}\left[E_k^{*}(s)-E_{k_1}(s)\right]ds\right) \quad (11)$$

For the rectangular potential barrier with height $U_0$ and width $\Delta$ the solution of the Schrödinger equation can be written as a plane waves $P_k \approx \exp(ik_xX)$, $\varphi_k = k_xX$, where the wave number $k_x$ are related to the energy via $k_x = \pm(2m(E-E_k))^{1/2}/\hbar$. Note that if the energy $E < E_k$ the wave number $k_x$ becomes imaginary and $P_k \approx \exp(-\kappa X), \kappa = \left(2mU_0/\hbar^2 - K_x^2\right)^{1/2}$ but $I_k = 0$ at $0 < x < \Delta$ because $P_k = P_k^*$. It is instructive to note here that the complete set of $K_x$ (4) corresponds to each value of coordinate X. Thus for semiconductor with depletion layer in the region of Schottky barrier distribution function can be represented in the form

$$n(X,K_x) = n(X)n_e(K_x)g(X,K_x)A^{-1}(X,V)T(X,K_x) \quad (12)$$

Here $T(X,K_x)$ is the barrier-transmission coefficient, A(X,V)- normalizing factor, determining from the equation

$$N=2S\sum_{X,K_x}n(X,K_x)=Sd_x\sum_X n(X), A(X,V)=\left(\frac{\hbar^2}{2\pi mkT}\right)^{1/2}\int_{-\infty}^{\infty}f(X,K_x)\exp\left(-\frac{\hbar^2K_x^2}{2mkT}\right)T(X,K_x)dK_x \quad (13)$$

In the neutral region $A(X,V)\approx 1, T(X,K_x)\approx 1$. The dependence of the normalizing factor $A(X,V)$ at the $\Delta > X > 0$ (Schottky barrier region, $\Delta$ is the width of barrier) on the applied voltage $V$ for simple rectangular barrier model is visualized in Fig.1.

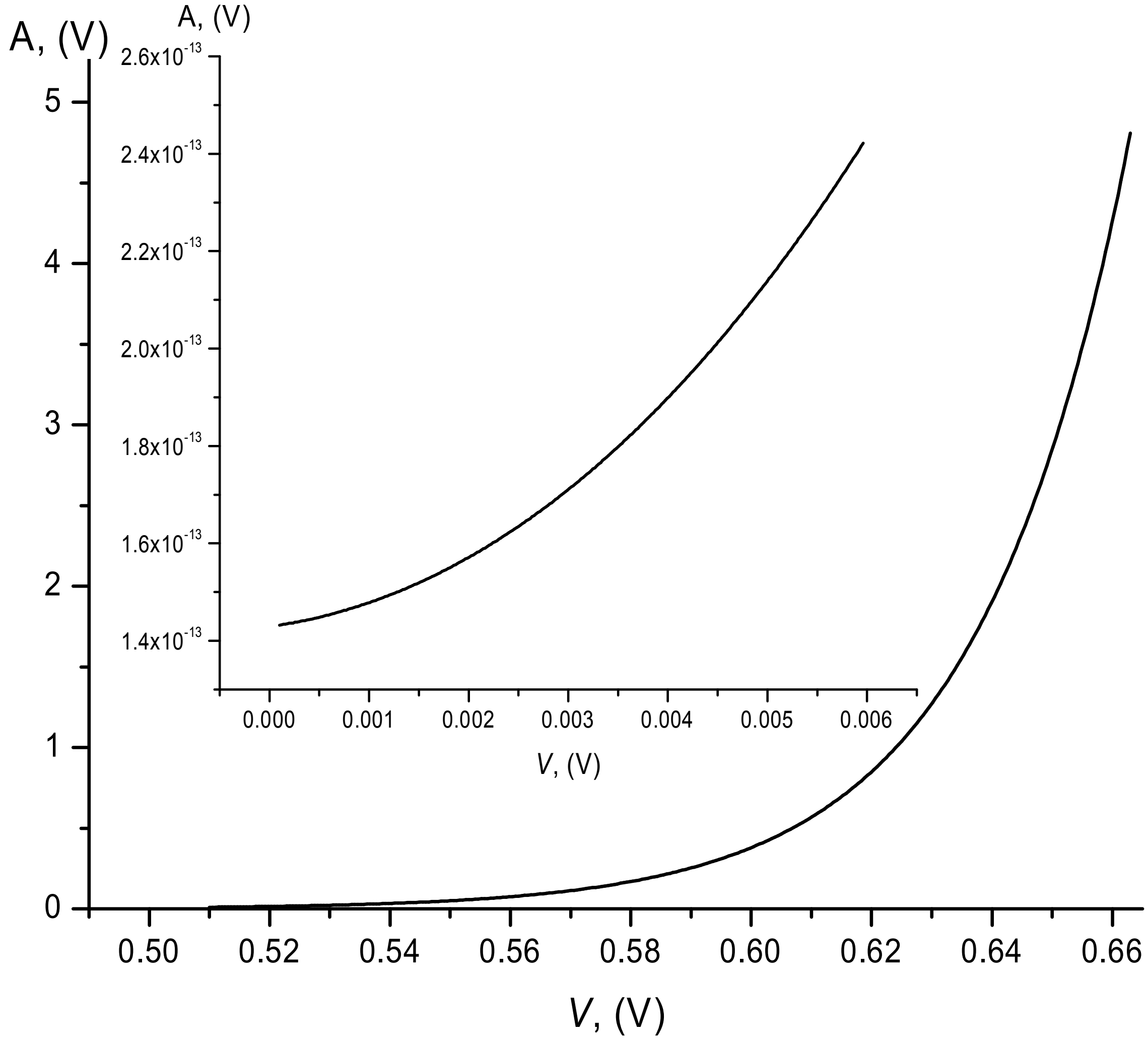

Fig.1. The computed normalizing factor $A(\Delta,V)$ is shown as function of applied voltage $V$ .

Equation (13) is solved for a model with $d = 2\mu m$ wide ZnSe semiconductor. For the calculation a temperature of 300 K, mobility $\mu_n = 0.053\,\text{cm}^2/\text{Vs}$, $\lambda(0)/\Delta = 10$, $N_D = 10^{21} m^{-3}$ and effective mass of electrons $m = 0.17 m_0$ are used.

Insert shows that the value of $A(\Delta, V)$ varies significantly from $1.42\cdot10^{-13}$ at $V = 0$ up to 5 at $V = 0.66\,\text{V}$, such that one obtains very incorrect estimates of concentration $n(x)$ if one ignores the variations of factor $A(\Delta, V)$ and takes $A(x,V) = 1$. Quantum transmission coefficient $T(X, K_x)$ can be calculated having applied the Wentzel–Kramers– Brillouin approximation ($E < U$) for depletion region

$$T(X,K_x) = \exp(-\frac{2}{\hbar}\int_0^X \sqrt{2m(U(x)-E)dx}), U(x) > E,\ T(X,K_x) = \exp\left(-\frac{X}{\lambda}\right),\ U(x) < E, \quad (14)$$

Here $U(x)$ - potential energy. As the simple approximation the probability of electron emission over Schottky barrier ($E > U$) is given by $T(X, K_x) = \exp(-X/\lambda)$, $\lambda$ is the mean free path of carriers.

It is common practice to consider the tunneling current as additional quantum correction to the classic current expressed in terms of the equation (1). In that case tunneling current is determined via the product of the quantum transmission coefficient and the distribution function. We determine current density by the next equation

$$j(X) = -q\frac{\hbar}{m}\int_{-\infty}^{\infty}\frac{dK_x}{2\pi}2K_x n(X, K_x) \quad (15)$$

where transmission coefficient is included into distribution function $n(X, K_x)$ (12) as multiplier. Equation (12) shows that tunneling effect is described along with thermionic emission, drift in electric field and diffusion in the united approach for which one has to know nothing else but the solution of difference equation for microscopic polarization $P_{kk'}(t)$ (9, 12). We only have to know distribution function $n(x, k_x)$ to calculate current density $j_x(X)$ (15). We will resolve the transport problem by means of distribution function $n(x, k_x)$ in the form (12).

### III. Numerical results and discussion.

Inserting $n(x,k_x)$ (12) into the equation for current density (15) and evaluating integral over $K_x$ we obtain for current density at the metal-semiconductor interface $X = 0$ the result

$$j(0) = \frac{1}{4}qn(\Delta)v_T A^{-1}(V_\Delta)F(V_\Delta), \quad v_T = \left(\frac{8}{\pi}\frac{\text{k}*\text{T}}{\text{m}}\right)^{1/2} \quad (16)$$

Here $V_\Delta = V - V_{d-\Delta}$, $V_{d-\Delta}$ is the parasitic voltage drop due to series resistance of neutral region. It is customary to assume that $V_{d-\Delta} \approx 0$. We have to note here that the density of electrons at the metal-semiconductor interface $n(0)$ is determined by the density of electrons before the barrier $n(\Delta)$ multiplied by the transfer coefficient in accordance with Eqs.(12,16). Function $F(V)$ is determined by the equation.

$$F(V) = \frac{\hbar^2}{mkT}\int_0^\infty \delta G(\Delta, k_x)dk_x, \delta G = G(V) - G(0),\ G(\Delta, k_x) = k_x g(\Delta, k_x)\exp\left(-\frac{\hbar^2 k_x^2}{2mkT}\right)T(\Delta, k_x) \quad (17)$$

. It can be shown that in the limit, where only thermionic current is taken into account one gets the well-known expression for function $F(V) = \exp(-qV_{bi}/kT)(\exp(qV/kT) - 1)$. Here $V_{bi}$ is the built-in potential.

Figure 2 depicts the main qualitative features of numerical results for function $F(V)$ for simple rectangular barrier model. In the limit where applied voltage tends to zero function $F(V) \to 0$ . The value of function $F(V)$ varies from zero up to 6.

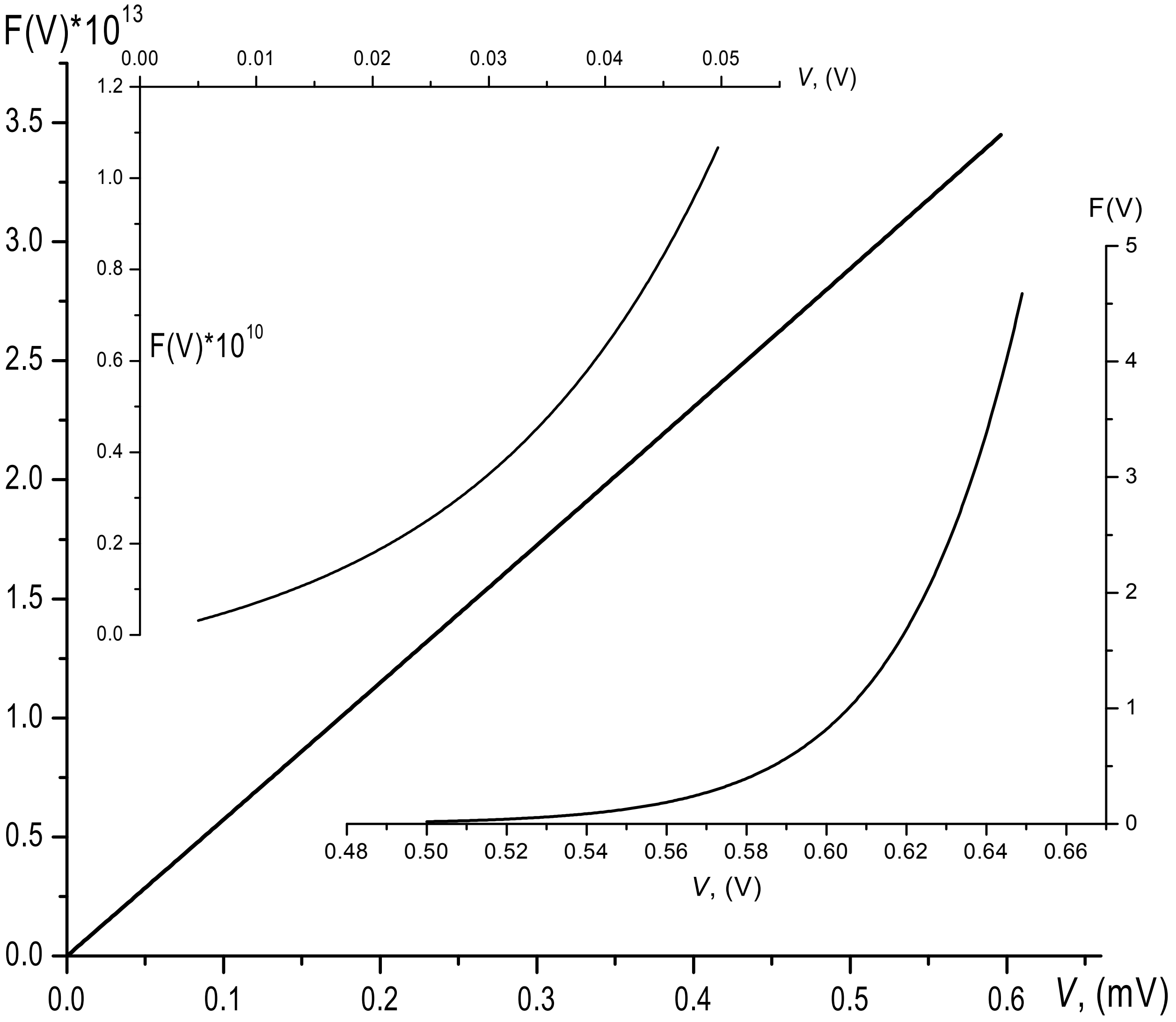

Fig.2. Calculated voltage-dependent function $F(V)$ for ZnSe MS structure. For the calculation the values of semiconductor's parameters given in the Fig.1 capture are used.

Total current on the semiconductor surface ( $x = 0$ ) is the sum of tunneling current and thermionic-emission current. Ratio of tunneling and thermionic-emission currents is determined by the equation

$$\frac{j_{therm}}{j_{tun}} = \int_k^\infty \delta G(\Delta, k_x) dk_x \Big/ \int_0^k \delta G(\Delta, k_x) dk_x, \quad k(V) = \sqrt{2mq(V_{bi} - V)} / \hbar \quad (18)$$

Fig.3 shows that at values of the parameter $\lambda(0)/4\Delta$ greater than 0.022 thermionic-emission current dominates. If $\lambda(0)/4\Delta$ =0.12 than the ratio $j_{therm} / j_{tun}$ is about $10^4$. But at. $\lambda(0)/4\Delta$ =0.012 the value of $j_{therm} / j_{tun}$ is about $10^{-4}$ The increasing of the barriers width $\Delta$ leads to the decreasing of the tunneling current $j_{tun}$ .

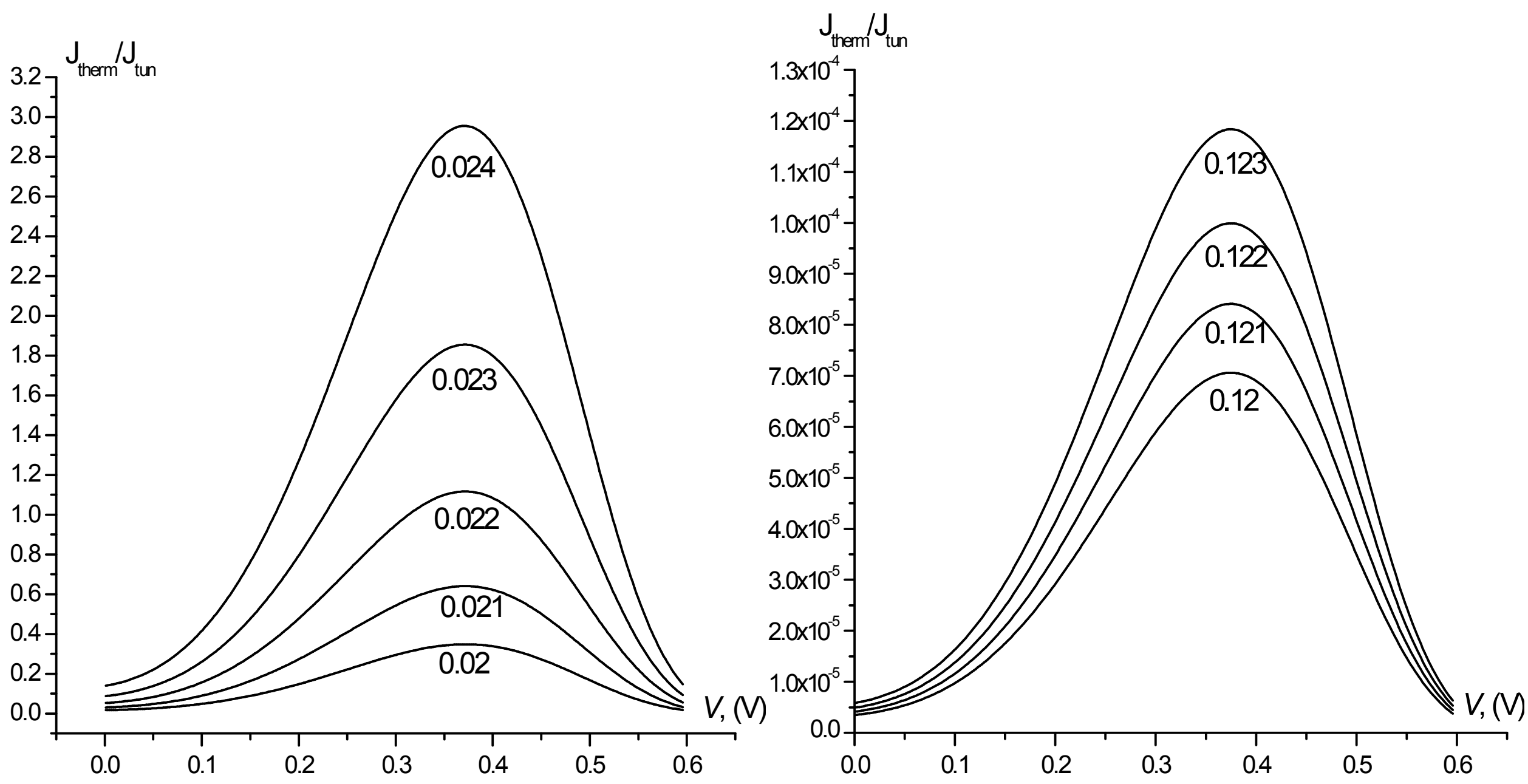

Fig.3. Relation $j_{therm} / j_{tun}$ at different values of the parameter $\lambda(0)/4\Delta$ indicated in figure.

Considering neutral region, where expression for current density (1) is valid, we obtain the dependence of concentration on coordinate in the form

$$j(x) = \mu_n kT \left\{ n(x) e^{-\frac{q\varphi(x)}{kT}} - n(x') e^{-\frac{q\varphi(x')}{kT}} \right\} / \int_{x'}^{x} e^{-\frac{q\varphi(x_1)}{kT}} dx_1 \quad (19)$$

Electric potential $\varphi(x)$ is determined by the distributions of electrons and shallow impurities according to the Poisson equation. In the neutral region potential $\varphi(x)$ is the linear function of x, accordingly

$$j(d) = q\mu_n E(d) n(d), (d-\Delta)E(d) = V_{d-\Delta} \quad (20)$$

For depletion region the equations (19, 20) are no longer valid. It is necessary that current density at the metal-semiconductor interface (16) satisfy the continuity condition. Equation of continuity $j(0) = j(d)$ can be easily resolved in the limit, where applied voltage $V \to 0$ and $j(0) \approx V_\Delta$, $j(d) = q\mu_n n(d) V_{d-\Delta} / (d-\Delta)$. Equation (19) at $x = d, x' = \Delta$ together with continuity condition $j(0) = j(d)$ in view of $V_\Delta = V - V_{d-\Delta}$ at arbitrary applied voltage implies the following result

$$j = \mu_n kT \frac{n(d)}{d-\Delta} \left( 1 + \frac{\pi\lambda(d)}{2(d-\Delta)} \frac{A(V)}{F(V)} \exp[(\beta_{bi} - \beta)\left(1 - \frac{\Delta}{W}\right)^2] \right)^{-1}, \beta = \frac{qV}{kT}, \beta_{bi} = \frac{qV_{bi}}{kT}, W = \left( \frac{2\varepsilon\varepsilon_0}{qN_d} [V_{bi} - V - \frac{kT}{q}] \right)^{1/2} \quad (21)$$

When the impurity concentration changes abruptly the potential drop is determined by the equation $\varphi(d) - \varphi(\Delta) = (V_{bi} - V)(1 - \Delta/W)^2$. One can see from equations (21) that the ideality factor in the zero approximation $n_0$ is given by the next formulae

$$n_0 = \left( 1 - \frac{\Delta(V,T)}{W(V)} \right)^{-2} \quad (22)$$

Effective width of Schottky barrier $\Delta$ is less than depletion region width $W$. Hence $n_0 \geq 1$. In the limit where $\Delta \to 0$, ideality factor tends to 1. Further analytical expression for the above equation (21) is difficult, and results can be obtained by numerical calculation.

Some of the parameters of theory expressed by the Eq.(21) $V_{bi}, \mu_n, m_n, d, N_D, T$ are well established, others such as relaxation time $\tau(d) = \mu_n m_n / q$, at the semiconductor surface x=d and $\tau(d)v_T = \lambda(d)$, $\pi\lambda(d)/2(d-\Delta)$ can be calculated. But effective width of Schottky barrier $\Delta$ and relaxation time $\tau(0)$ can be only estimated.

Figure 4 displays the I vs forward voltage V curves computed by using different values of parameter $\lambda(0)/\Delta$, where $\lambda(0)=\tau(0)v_T$ is the mean free path of electrons at the interface $x=0$. Calculated values of typical current-voltage characteristics show that in the region of small electric field $V<0.01$V current reaches its maximum if parameter $\lambda(0)/4\Delta>1.4$. At high voltage $V>0.1$V the ratio $A(0,V)/F(0,V)$ do not depend on parameter $\lambda(0)/\Delta$, therefore all curves showing in the Fig.4 merge into one.

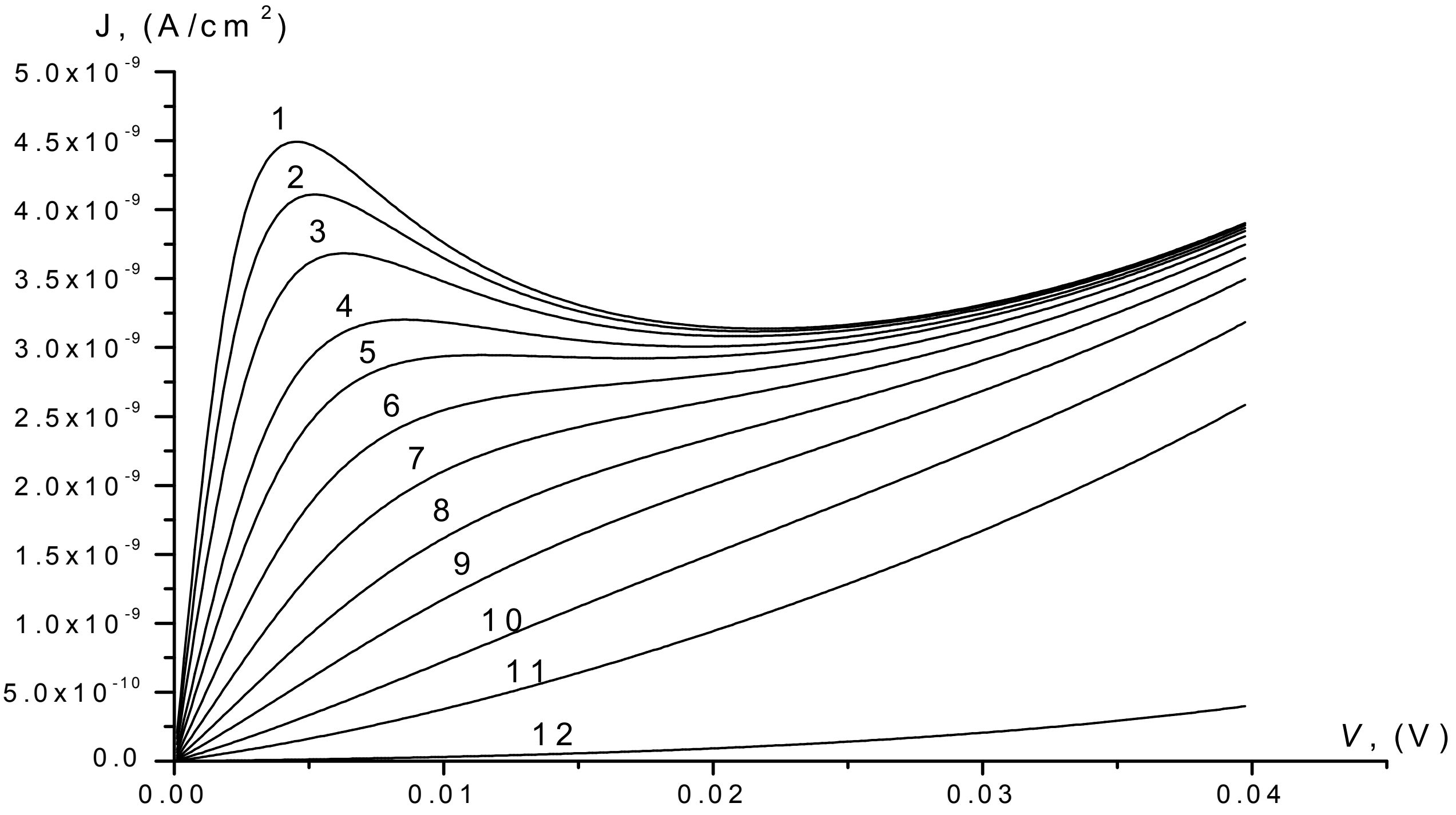

Fig.4. Theoretical I-V characteristics of different MS (ZnSe) structures for different values of parameter $\lambda(0)/4\Delta$: 1-$\lambda(0)/4\Delta$ =5, 2-4, 3-3, 4-2, 5-1.5, 6-1, 7-0.65, 8-0.4, 9-0.24, 10-0.12, 11- 0.06. 12-0.03. The curves are obtained assuming simple rectangular barrier model with $\pi\lambda(d)/2(d-\Delta)=0.008$

But at a fixed value of $\lambda(0)/\Delta$ and different $\lambda(d)/d$ a curves $j(V)$ at $V>0.1$V depend on parameter $\lambda(d)/d$ according to Fig.5. If applied voltage is in the range 0.5V $> V >$ 0.1V the current density obey the exponential dependence on $V$ due to the Boltzmann distribution and ideality factor at 0.5V $>V>$ 0.1V is equal approximately to one( $n\cong 1$) at $\Delta/W \ll 1$.

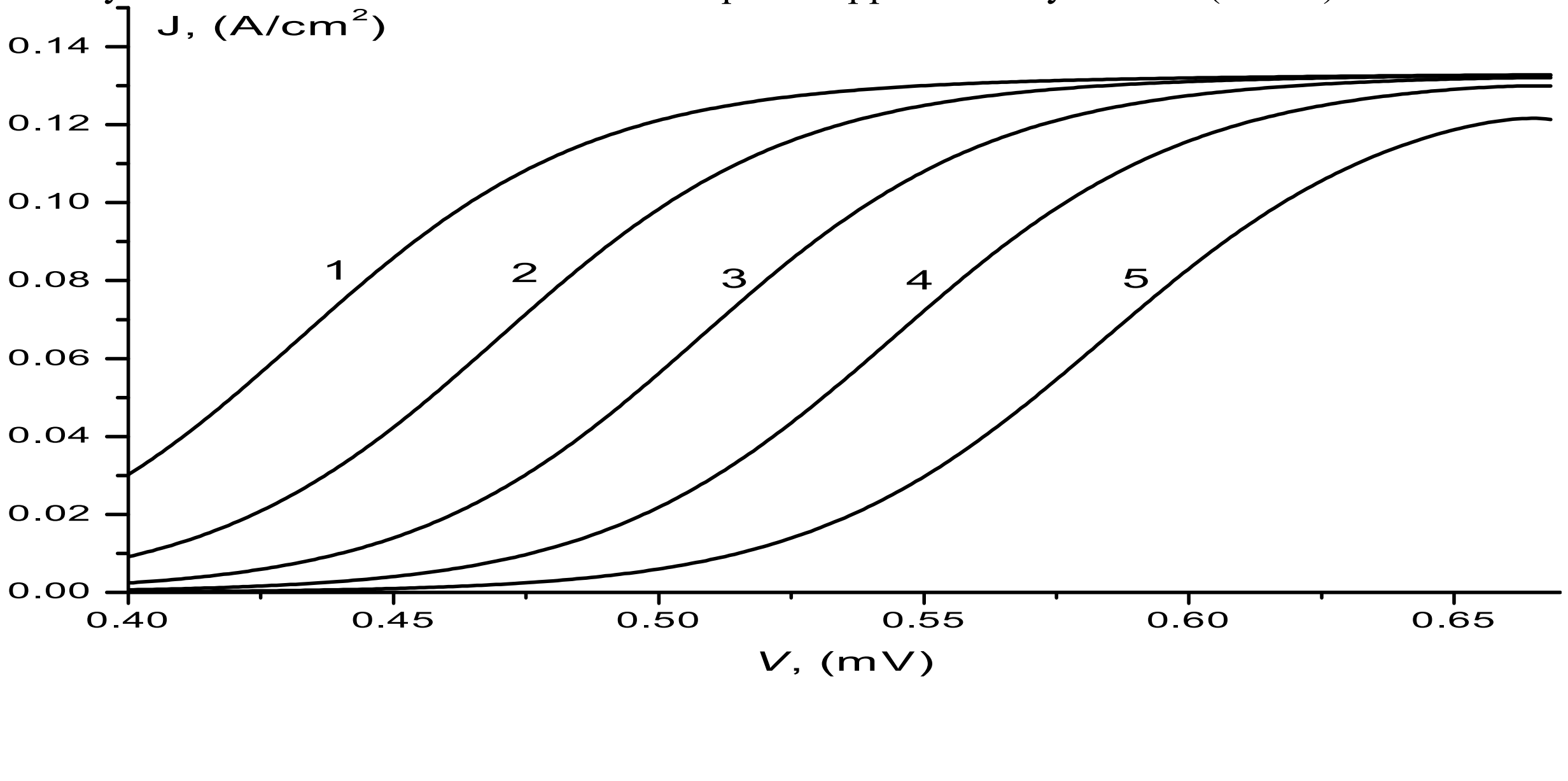

Fig.5. Same as Fig.4, but in the high electric field $V > 0.4$V for $\lambda(0)/4\Delta = 0.1$. and different values of parameter $\pi\lambda(d)/2d$ : 1-$\pi\lambda(d)/2d$=0.0004, 2-0.0016, 3-0.0064, 4-0.024, 5-0.1

At high bias current density behavior (Fig.5) definitely departs from what is observed at low voltage $V$. As shown in figure 5 at high bias exponential current voltage dependence is replaced by a power law. It is observed at $V$ >0.45 V for $\pi\lambda(d)/2d$ =0.0004 and at $V$ >0.60 V for $\pi\lambda(d)/2d$ =0.024.

Fig.6 a and b are a plots of ideality factor $n(V) = \frac{q}{kT} dV/d\ln j(V)$ versus applied voltage $V$. As it is clear from figures 6 a and b the magnitude of $n(V)$ increases with increasing of $V$ at small $V$. Then at 0.4V > $V$ > 0.1V the value of $n(V)$ reaches $n \cong 1$,($n_0$=1) Fig.6a. In the region of high voltage $V>0.4$V the ideality factor $n(V)$ increases. Hence, I-V characteristics acquire an ohmic nature. This effect is visualized in Fig.6a and b.

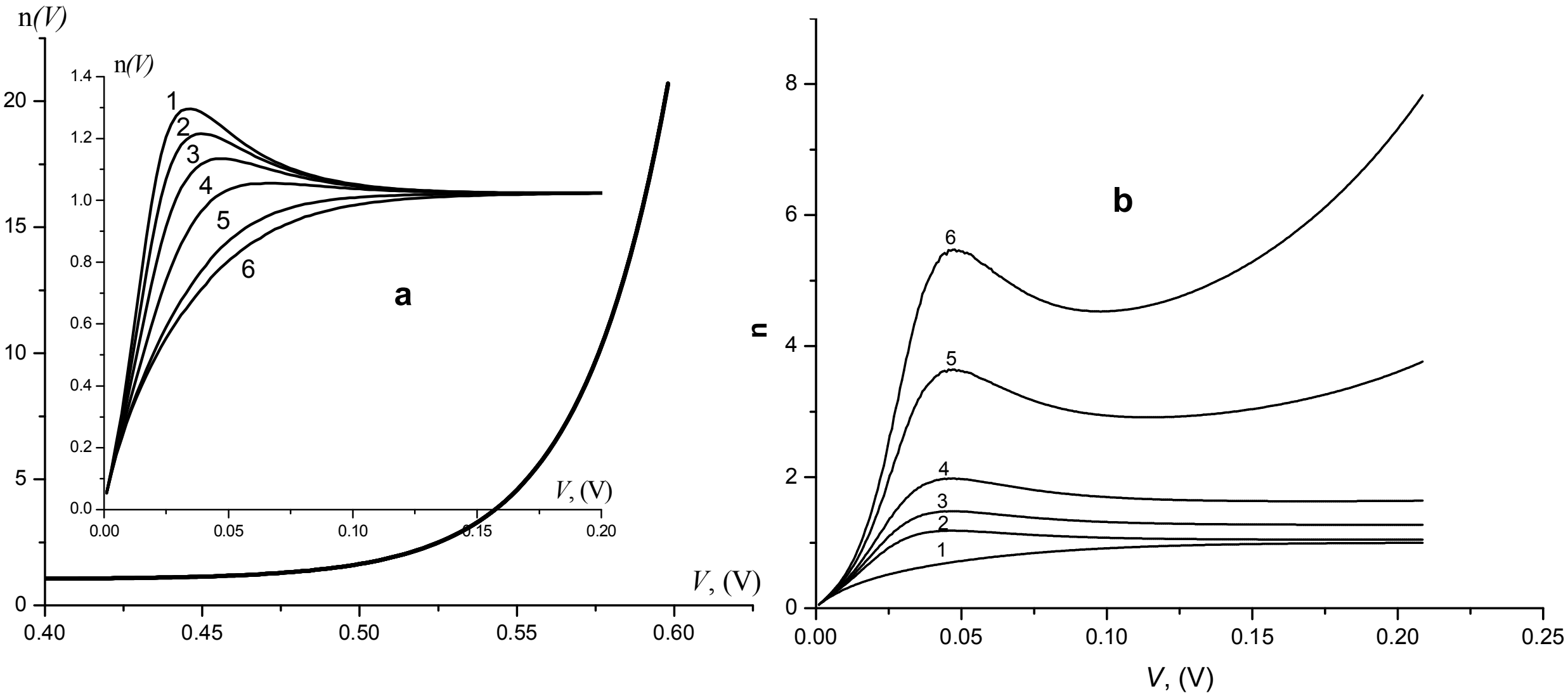

Fig.6. Ideality factor $n(V)$ calculated by using of Eq.(21) a) at different values of $\lambda(0)/4\Delta$, 1-$\lambda(0)/4\Delta$=0.5, 2-0.25, 3-0.05, 4-0.02, 5-0.005, 6-0.0001 and $n_0 = 1$, $\pi\lambda(d)/2d = 0.008$, b) at different $\Delta/W$ = 0.001-1,0.01-2,0.1-3,0.2-4,0.35-5, 0.4-6 for $\lambda(0)/4\Delta = 0.1$.

The value of $V$ at which I-V characteristics acquire an ohmic nature is shifted to the lower bias $V$ as parameter $\pi\lambda(d)/2d$ decreases. Current -voltage calculations at different temperatures are shown in Fig.7

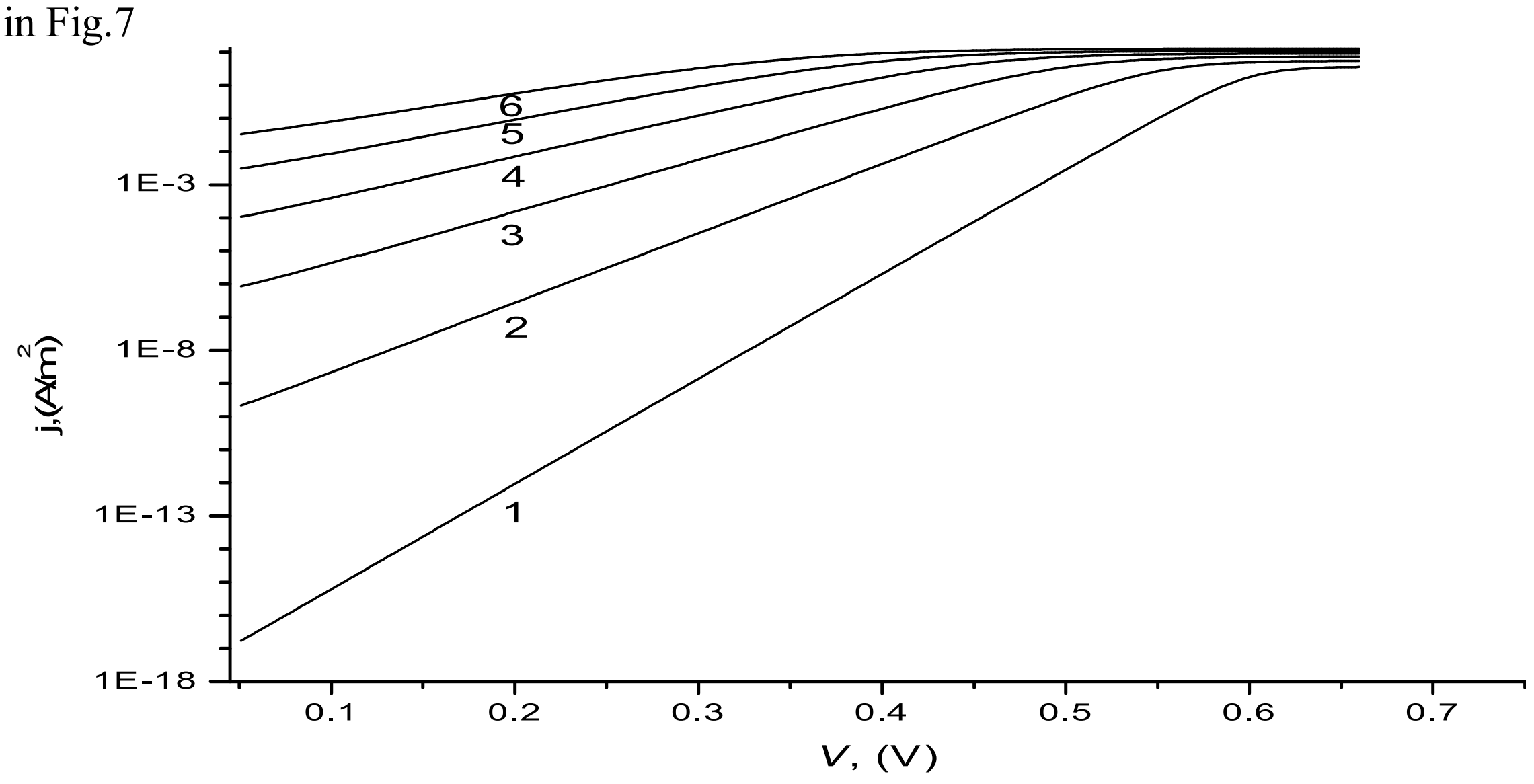

Fig.7 The variation of I-V forward characteristics of ZnSe MS Schottky junction with temperatures. 1- T=100K, 2-150K, 3-200K, 4-250K, 5-300K, 6-350K at $\Delta/W$=0.2, $N_D = 10^{21} m^{-3}$, $\lambda(0)/4\Delta$ =0.1 and $\pi\lambda(d)/2d$ =0.008.

. It can be concluded that it is possible to phenomenologically adjust the parameters of the theory, Eq.(21), to reproduce the main signatures of results for the I-V characteristics of MS structures at different temperatures.

Examples of a measured I-V characteristics of MS structures and the comparison with theory are demonstrated in Figure 8 [8,9].

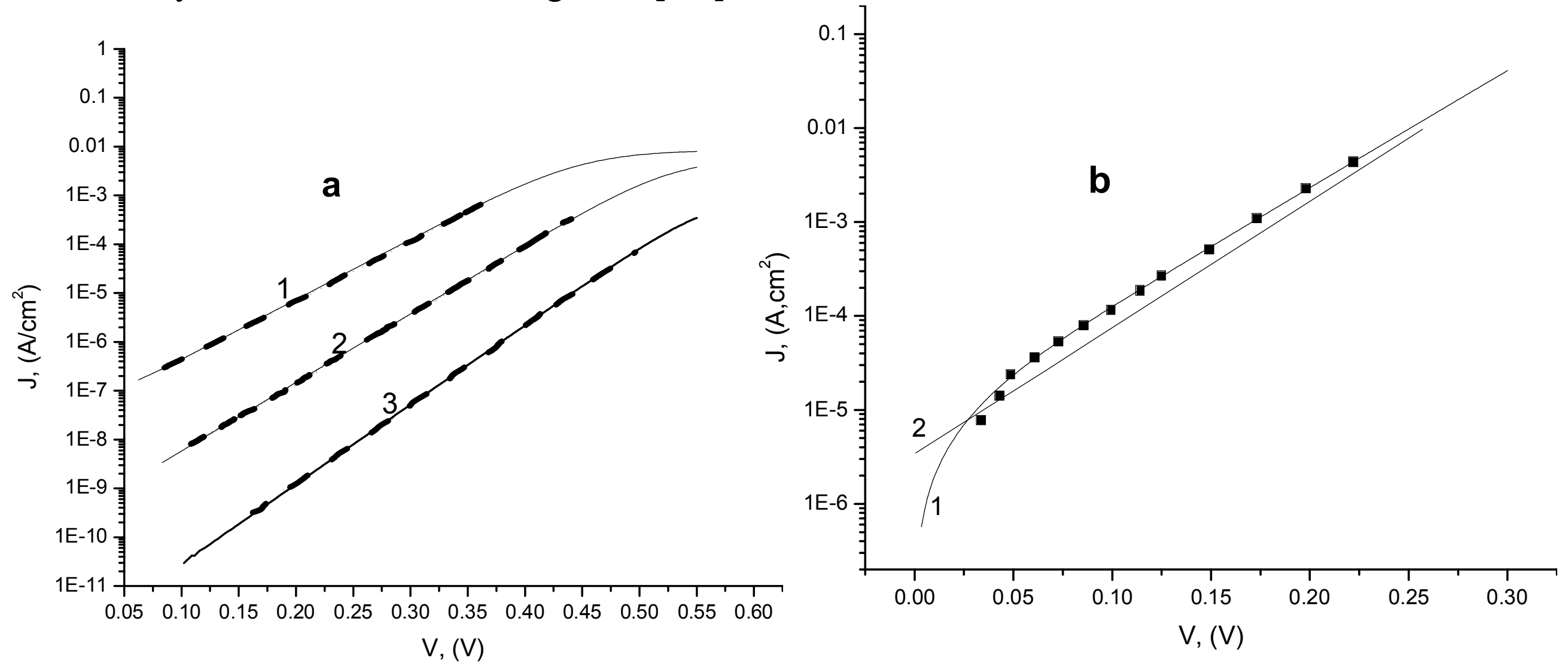

Fig.8. Experimental (symbols) and theoretical (lines) forward I-V characteristics of different MS structures. (a) An Au-GaAs diode showing a comparison between calculated, Eq.(21), and experimental curves from reference [8] for different temperatures.1-373K, $\Delta/W$=0.039,2-313K, $\Delta/W$=0.0542,3-253K,$\Delta/W$=0.0851,. b) An I-V characteristic of an Au-Si diode, 1-theoretical line is obtained with Eq.(21) assuming $\Delta/W$=0.15, $\lambda(0)/4\Delta$ =0.09. 2- theoretical curve from [2,9], symbols-experiment [2,9].

The I-V forward characteristics of an Au-GaAs diode are shown in Fig.8(a) for different temperatures. It is well known that ideality factor n depends on the temperature [2,10]. For lower temperature n increases. This effect is visualized in Fig.8(a). At T=373 we have $\Delta/W$=0.039, $n_0$ =1.083, but at T=253 parameter $\Delta/W$ = 0.0851 and $n_0$ =1.19. The dependence of parameter $\Delta/W$ on temperature is due to the Maxwell's distribution of electrons. The effective width $\Delta$ is increased where temperature is decreased due to the decreasing of numbers of high-energy electrons. For higher temperature the effective width $\Delta$ of Schottky barrier decreases since high-energy electrons tunnel through thin potential barrier.

Theoretical and experimental values of current-voltage characteristics for Au-Si barrier at T=296K, $N_d$ =$10^{24}$ m$^{-3}$ are shown in the Fig 8(b). The main qualitative and quantitative features of experimental results, demonstrated in Fig.8 are in very good agreement with the developed theory, represented by Eq.(21).

## IV. Conclusions.

This article presents the microscopic many-body transport approach for electronic properties of spatially inhomogeneous systems which is developed at the quantum-mechanical level by means of plane wavelets second quantization representation. We have used the set of ket vectors Eq.(4) to introduce the positively definite distribution function which describes inhomogeneous system by numbers of particles at quantized positions with quantized momenta. We focus on the proof that distribution function can be represented by Eq.(12) which describes tunneling effect along with thermionic emission, drift in electric field and diffusion in the united approach. The example of distribution function Eq.(12) used in this article is not more than a one case where solution of kinetic equations (6-11) can be evaluated for simple rectangular barrier model. Quantum mechanical theory Eqs.(15-21) predicts that the current at the metal-semiconductor interface depends on the density of electrons before the barrier $n(\Delta)$. This is the reason why the ideality factor in the zero approximation $n_0$ is determined by the potential drop $\varphi(d)-\varphi(\Delta)$. The theoretical predictions for the dependence of the ideality factor $n_0 = (1 - \Delta(V,T)/W(V))^{-2}$ on the temperature agree with experimental data.

Hence we can conclude that the resulting approach simplifies the calculation of the I-V characteristics and should be useful for preliminary design.